\documentclass[12pt]{article}
\usepackage{hyperref}
\usepackage[compress]{cite}
\usepackage{lscape}
\usepackage{amsmath}
\usepackage{color}
\usepackage{graphicx}
\usepackage{amssymb}
\usepackage{varioref}
\usepackage[left=2cm,right=2cm,top=2cm,bottom=2.5cm]{geometry}
\labelformat{section}{Section #1} 
\labelformat{subsection}{Section #1} 
\labelformat{subsubsection}{Section #1}
\labelformat{subsubsubsection}{Section #1}
\labelformat{equation}{Eq.~(#1)} 
\labelformat{figure}{Fig.~#1} 
\labelformat{subfigure}{Fig.~\thefigure#1} 
\labelformat{table}{Tab.~#1} 
\labelformat{appendix}{Appendix #1}
\renewcommand{\title}[1]{%
    \bigskip%
    \begin{center}%
    \Large\bf #1%
    \end{center}%
    \vskip .2in}
\renewcommand{\author}[1]{%
    {\begin{center}
    #1
    \end{center}}}
\newcommand{\address}[1]{\vspace{-1.7em}\vspace{0pt}
    {\begin{center}
    \it #1
    \end{center}}}

\begin{document}


\title{Cosmological implications of shift symmetric Galileon field}

\author
{Rabin Banerjee \footnote{rabin@bose.res.in}$~^{\rm a}$,
Sumanta Chakraborty \footnote{sumantac.physics@gmail.com}$~^{\rm b}$,
Arpita Mitra \footnote{arpita12t@bose.res.in}$~^{\rm a}$,
Pradip Mukherjee \footnote{mukhpradip@gmail.com}$~^{\rm c}$}\vspace{0.5em}

\address{$^{\rm a}$ S. N. Bose National Centre for Basic Sciences\\
JD Block, Sector III, Salt Lake City, Kolkata 700 098, India. }

\address{$^{\rm b}$ Department of Theoretical Physics\\
Indian Association for the Cultivation of Science, Kolkata 700032, India.}

\address{$^{\rm c}$ Department of Physics, Barasat Government College\\
10, KNC Road, Barasat, Kolkata 700124, India.}

\begin{abstract}
A ghost-free metric formulation of the recently proposed covariant Galileon model \cite{RPgal} which retains the internal shift symmetry has been constructed. This presents a new result because the covariant Galileon models introduced so far in the literature lacks the full Galileon symmetry. We demonstrate that the same general procedure is applicable to all the Galileon interaction terms  and the resulting ghost-free Galileon action respecting internal shift symmetry cannot predict the late time acceleration of the universe. However, weakly breaking the shift symmetry in the covariant Galileon Lagrangian, allows us to include an appropriate potential term which can explain late time acceleration, in accordance with recent findings in the literature.
\end{abstract}
\section{Introduction}

Recent cosmological observations \cite{obe19,obe18,obe3,obe4,obe1} indicate late-time acceleration of the observable universe, a phenomenon that can not be explained in the standard cosmological scenarios. Various theoretical attempts have been undertaken to confront this observational fact. Though most observational evidences favour the $\Lambda$CDM model, it seemingly have certain conceptual difficulties, e.g., the fine tuning and coincidence problems owing to the absence of any dynamics of the cosmological constant. Furthermore, the huge difference of energy scale, $\mathcal{O}(10^{-127})$, between the vacuum energy of quantum theory and that of cosmological constant is of serious discomfort. However there have also been claims that the vacuum energy should not gravitate through a stress tensor and hence the fine tuning problem never arises, or the gravitational dynamics must be impermeable to constant shifts in the matter energy momentum tensor \cite{Dadhich:2010ca,Padmanabhan:2014nca,Padmanabhan:2017qvh}. Alternatively, one conjectures an exotic dark energy (DE) fluid (for a set of reviews, see \cite{DE,Paddy_1,DE_Review_Sami}) described by a scalar field with a suitable dynamics of its own.

Scalar fields have long been favourite in dark energy model building. One such scalar field model comes with a canonical kinetic term and in addition appropriate self interactions, known as the quintessence scenario. Interestingly, such models give rise to negative pressure, a key element in explaining the late-time acceleration, leading to an equation of state (EoS) parameter $\omega$, bounded within the range $-1 < \omega <-(1/3)$.  However in the past, observational data (see e.g., \cite{obe1,obs_data_Phamtom_crossing} have even allowed $\omega$ to cross the de-Sitter limit $\omega =-1$. The recent Planck observations \cite{Planck_1} as well have produced various combined data sets on the EoS parameter, some of which (e.g., $95 \%$; Planck+WMAP+SNLS and $95 \%$; Planck+WMAP+${\rm H}_{0}$) prefer certain phantom evolution of the DE fluid with $\omega < -1$. Scalar field models with wrong sign of the kinetic energy (phantom, ghost condensate model etc.) term were proposed to include $\omega < -1$ regime but these 
models have almost insurmountable stability problems. Scalar fields with non minimal coupling, derivative coupling as well as multiple scalar fields have been introduced and discussed zealously but still the origin of dark energy (or, for that matter late time accelerating phase of the universe) remains an open problem in the current cosmological research (see also \cite{Chakraborty:2014fia}).

A class apart in the realm of scalar field theories, is the Galileon model \cite{N}. It was obtained by taking the decoupling limit of the DGP model \cite{D} leading to an effective theory for a scalar field and argued to describe the scalar sector of the original model \cite{LP,NR}. Galileon theories are interesting in several ways, in particular they are characterized by two properties:
\begin{enumerate}

\item Despite being higher derivative theories, they are free of Ostogradsky-type ghosts since they have no more than second order equations of motion \cite{Galileon_2nd_order_eqm_1,Galileon_2nd_order_eqm_2,Galileon_2nd_order_eqm_3}.

\item They are invariant under a nonlinear symmetry transformation of the scalar field $\pi$,
\begin{eqnarray}
\pi \to \pi + c + b_\mu x^\mu \label{shift}
\end{eqnarray}
\end{enumerate} 
where $c$ and $b_\mu$ are constants. A direct consequence of this invariance is that the coefficients of the leading Galileon interactions are, perturbatively, not renormalized \cite{hamed}. Due to these non-trivial features, it becomes very inviting to consider it as a model for dark energy.

In recent times a lot of interest has been focused on the dark energy models with Galileon scalar field 
\cite{Galilean_1,ref3,Galilean_2,Galilean_3} due to the absence of negative energy instability. However, in the original Galileon scalar field cosmology \cite{Galilean_1}, the Galilean symmetry $\partial_{\mu} \pi \to \partial_{\mu} \pi + b_{\mu}$ is essentially broken when gravity is introduced. Indeed the lifting of the original flat space Galileon symmetry to a curved background was not clear. A covariant formulation of the Galileon model was obtained in \cite{Galileon_2nd_order_eqm_1} where only the shift symmetry $\pi \to \pi+c$ could be preserved. In fact the general consensus is that it is impossible to couple Galileons to gravity while retaining \ref{shift} or its curved space extension \cite{P}. 
It is evident that applications of the Galilean model to cosmology would require a covariant theory of the Galileon field. Several authors have contributed in this field and it was correctly concluded that in a curved spacetime a minimal or even non minimal coupling of the Galileon scalar field breaks the shift symmetry, mainly due to the fact that a constant shift vector is inconsistent with the invariance of the Galileon Lagrangian in curved spacetime. However even in that scenario the flat spacetime limit enjoys the full Galileon symmetry. One wonders whether the symmetry of the flat space model can be extended in such a way that the corresponding curved space model retains the Galileon symmetry in some form. Of course the curved space generalization must give the Galileon model in the flat spacetime limit.

The above requirement was taken as the point of departure in \cite{RPgal} and a localized form of the transformation in \ref{shift} was assumed. The powerful apparatus of localizing the symmetry, introduced for Poincar\'{e} symmetry by Utiyama \cite{U,K,S,W1,BH,carroll} and recently generalized to different contexts \cite{MM1,MM2,MM3,BM4,MMM2,BGM}, was then applied in \cite{RPgal} to couple the Galileon model with gravity in curved spacetime which retains both the properties, namely, invariance under covariant Galileon symmetry and absence of Ostrogradsky ghosts. The covariant Galileon model thus constructed involves 
auxiliary fields corresponding to the localization of Poincar\'{e} (spacetime) and Galileon (internal) symmetries. The gauge fields corresponding to Galileon symmetries can be clubbed into a single vector field $F_\mu$. Now, the model in \cite{RPgal} was formulated in the first order (tetrad) formalism whereas for cosmological applications a metric formulation is more suitable. It is only natural to inquire how the metric formulation of our model behaves in comparison with the covariant Galileon theories introduced earlier in the literature \cite{Galilean_1,ref3,Galilean_2,Galilean_3,Galileon_2nd_order_eqm_1}. The crucial difference corresponds to the fact that our model preserves the full internal shift symmetry even in curved spacetime, while those in the literature do not. This is an important point, which we will derive later in this work. In addition, the model derived in this work can be applied to various scenarios, e.g., in cosmology and it would be interesting to explore whether it can give rise to late time acceleration. This becomes all the more interesting, for it turns out that in order to have a de Sitter vacuum  the covariant Galileon models considered in the literature are not sufficient \cite{P}, one needs to modify it further. The modifications are generally achieved by introducing arbitrary functions of the kinetic term, which helps to achieve a de Sitter vacuum, important for both inflationary paradigm as well as to bring late time cosmic acceleration in the canvas. In lieu of all these new ingredients one has to pay a hefty price, the Galileon symmetry is broken even at the level of flat spacetime. Thus one cannot make contact with the original Galileon model in flat spacetime, which has the full internal shift symmetry. This unpleasant feature is absent in our analysis as the original Galileon model in flat spacetime is trivially achieved.

In the standard literatures one generally invokes additional scales in the problem and try to argue that the energy scale at which Galileon symmetry is broken is much higher than the energy scale in the original problem. This suppresses all those terms leading to a \emph{weakly broken} Galileon theory \cite{P}. Cosmology with weakly broken Galileon symmetry produced important results, in particular the model gives rise to late time acceleration of the universe \cite{AASen11} and is consistent with the solar system tests of gravity through the Vainshtein mechanism \cite{AASen12,Bhattacharya:2016naa}. The relaxation of the shift symmetry allows one to introduce self interaction potentials as well. However, Galileon cosmology, with its full symmetry has not been studied earlier in detail. 

The paper is organized as follows: In \ref{Section_02} the metric formulation of the covariant Galileon model, first introduced in \cite{RPgal} has been explicitly worked out. The constraint on the gauge field occurring due to the localization of Galileon symmetry has been utilized to write down the action in terms of a new scalar field. This equivalent theory also exhibits the shift symmetry. The gravitational action is taken to be the usual Einstein-Hilbert action. In \ref{Section_03} the dynamics of the Galileon field obtained in \ref{Section_02} has been discussed. In \ref{Section_04} we have studied the cosmological implications of our model, in particular the value of the equation of state (EoS) parameter is shown to lie between $\frac{1}{5}\leq \omega \leq 1$, thereby demonstrating that there is no negative pressure. Thus the covariant Galileon model with localized shift symmetry does \emph{not} lead to late time acceleration of the Universe. It may be recalled that late time acceleration of the Universe is predicted by Galileon models which do not have shift symmetry. Following which we show by weakly breaking the shift symmetry of our model that late time acceleration is reproduced. This is done by the reconstruction method recently introduced in the literature. We end with a discussion on our results.
\section{Metric formulation of the covariant Galileon model}\label{Section_02}

In flat spacetime the most general scalar field model invariant under internal Galileon transformations as in \ref{shift} is given by three more interaction terms in addition to the canonical kinetic term in four space-time dimensions, 
\begin{eqnarray}
\label{gallag}
\mathcal{L}=(\partial\pi)^2 +\sum_{I=3}^5\frac{c_a}{\Lambda_3^{3(I-2)}}\mathcal{L}_i ~.
\end{eqnarray}
Here $c_{a}$'s are arbitrary constants and $\Lambda _{3}^{-1}$ is an additional length scale associated with this problem. The three interaction terms are respectively given by, 
\begin{align}
\label{gal1}
\mathcal{L}_3 &= (\partial\pi)^2~[\Pi]~, \\
\label{gal2}
\mathcal{L}_4 &= (\partial\pi)^2~ \big([\Pi]^2-
[\Pi^2]\big )~,\\
\label{gal3}
\mathcal{L}_5 &= (\partial\pi)^2~\big([\Pi]^3-3[\Pi][\Pi^2]+ 2[\Pi^3]\big )~,
\end{align}
where we denote $[\Pi]\equiv \Box\pi,~[\Pi^2]\equiv \partial^\mu\partial_\nu\pi\partial^\nu\partial_\mu\pi$, etc. In addition to being invariant under \ref{shift} Galileon theories share another special property: the associated scalar field equations are second order, both in time and space, despite the higher-derivative interactions \footnote{Actually the theory has higher powers of second as well as first derivative terms, which are equivalent to higher derivative terms \'{a} la total divergences.} in \ref{gal1} to \ref{gal3}. This ensures that there are no ghosts hidden in the Galileon field $\pi$.

In \cite{RPgal} we have localized this internal symmetry and provided the prescription to couple a generic field theory described by the following action in flat spacetime,
\begin{equation}
\int d^4x{\cal{L}}\left(\pi,\partial_\mu \pi, \partial_\mu\partial_{\nu}\pi
\right)\label{action}
\end{equation}
with gravity. The action in \ref{action} is Poincar\'{e} symmetric and is quasi-invariant under \ref{shift}. In other words, the action is invariant under internal Galileon transformations up to a boundary term. The localization method results in covariant Galileon model, again quasi-invariant under the transformation as in \ref{shift}, where $b^\mu$ is now assumed to be covariantly conserved. Covariant derivatives are defined which transform canonically. Replacing the ordinary derivatives by the covariant derivatives and after a correction to the measure, we get the required action for Galileon field in curved spacetime. The new fields which appear in the covariant derivatives are identified with the tetrad and the spin connection. Thereby, we get covariant Galileon model in curved spacetime. Note that in addition to the usual gauge fields arising due to localization of Poincar\'{e} symmetry, new gauge fields are also introduced corresponding to the Galileon symmetry. Effectively, we  get a single vector 
field on which a constraint is imposed due to the covariant constraint on the Galileon shift vector. The importance of this constraint will be elucidated in the following discussion. 

Looking back at the Galileon model in the flat space we see that the essence of the Galileon properties are captured by $\mathcal{L}_3$ alone. Thus to illustrate the basic features, for the moment being, we will drop $\mathcal{L}_4$ and $\mathcal{L}_5$ in \ref{action} and shall concentrate on the Galileon Lagrangian $\mathcal{L}_3$. Following the procedure outlined above \cite{RPgal} the action for Galileon field on curved background takes the following form,
\begin{align}
S_{\rm g}=\int d^4x \frac{1}{\Sigma}\left[-\frac{1}{2}\left(\eta^{\gamma\delta}{\bar{{\cal{D}}}}_{\gamma}\pi{\bar{{\cal{D}}}}_{\delta}\pi\right)+ \alpha\left(\eta^{\alpha\beta}{\cal{D}}_{\alpha}{\bar{{\cal{D}}}}_{\beta}\pi\right)\left(\eta^{\gamma\delta}{\bar{{\cal{D}}}}_{\gamma}\pi{\bar{{\cal{D}}}}_{\delta}\pi\right)\right]
\label{la}
\end{align}
where $\alpha$ is a constant parameter of dimension $(\textrm{length})^{3}$. The covariant derivatives present in \ref{la} are defined as follows: under Poincar\'{e} symmetry transformations alone the form of covariant derivative of a generic field $\phi$ is well known
\begin{align}
{\cal{D}}_{\alpha}\pi &=\Sigma_{\alpha}{}^\mu D_\mu \phi, D_\mu \phi = \partial_\mu \phi + \frac{1}{2}B_\mu{}^{\alpha\beta}\sigma_{\alpha\beta}\phi\notag\\
\label{covp}
\end{align}
where $B_\mu{}^{\alpha\beta}, \sigma_{\alpha\beta}$ are spin coefficients and spin matrices respectively and $\Sigma_{\alpha}{}^\mu$ are the tetrads. This is extended to include symmetry under Galileon shift. The complete expression of the covariant derivative of the Galileon field $\pi$ is
\begin{align}
{\bar{{\cal{D}}}}_{\alpha}\pi &= \Sigma_{\alpha}{}^\mu{\bar{ D}}_\mu \pi;{\bar{ D}}_\mu \pi =\partial_\mu\pi + F_\mu;  \left(F_\mu = A_\mu + x_\mu D\right)\label{cov}
\end{align}
In \ref{cov} the covariant derivative corresponds to both the localization of Poincar\'{e} and Galileon symmetries. The new fields $A_{\mu}$ and $D$ which appear due to localization of Galileon symmetries are clubbed into the single vector field $F_\mu$. 

Using the definitions of covariant derivatives coined earlier, from \ref{la} we obtain,
\begin{equation}
S_{\rm g}=\int d^4x \frac{1}{\Sigma}\left[ -\frac{1}{2}\left(\eta^{\gamma\delta}\Sigma_{\gamma}{}^\rho\Sigma_{\delta}{}^\sigma{\bar{D}}_{\rho}\pi{\bar{D}}_{\sigma}\pi\right)+\alpha\left(\eta^{\alpha\beta}\Sigma_{\alpha}{}^\mu D_{\mu}(\Sigma_{\beta}{}^\nu {\bar{D}}_{\nu}\pi)\right)\left(\eta^{\gamma\delta}\Sigma_{\gamma}{}^\rho\Sigma_{\delta}{}^\sigma{\bar{D}}_{\rho}\pi{\bar{D}}_{\sigma}\pi\right)\right]\label{la1}
\end{equation}
This particular form of the action for Galileon field is entirely in the tetrad formalism. However, it can be further modified to metric formulation. Using the vielbein postulate
\begin{equation}
D_{\mu}\Sigma_{\beta}{}^\nu =-\Gamma^{\nu}_{\rho\mu}\Sigma_{\beta}{}^\rho
\label{vp}
\end{equation}
we write \ref{la1} as,
\begin{align}\label{la2}
S_{\rm g}=\int d^4x\frac{1}{\Sigma}\Big[ -\frac{1}{2}\left(\eta^{\gamma\delta}\Sigma_{\gamma}{}^\rho\Sigma_{\delta}{}^\sigma\partial_{\rho}\pi'\partial_{\sigma}\pi'\right)&+\alpha\left(-\eta^{\alpha\beta}\Sigma_{\alpha}{}^\mu \Gamma^{\nu}_{\rho\mu}\Sigma_{\beta}{}^\rho\partial_{\nu}\pi'+\eta^{\alpha\beta}\Sigma_{\alpha}{}^\mu\Sigma_{\beta}{}^\nu \partial_{\mu}\partial_{\nu}\pi'\right)
\nonumber
\\
&\times\left(\eta^{\gamma\delta}\Sigma_{\gamma}{}^\rho\Sigma_{\delta}{}^\sigma\partial_{\rho}\pi'\partial_{\sigma}\pi'\right)\Big]
\end{align}
In the above we have used another fact that the vector field $F_{\mu}$ introduced in the covariant derivatives satisfies the following constraint,
\begin{equation}
D_{\mu} F_{\nu}-D_{\nu} F_{\mu}= \partial_\mu F_\nu -\partial_\nu F_\mu = 0;\qquad F_{\mu}=\partial_{\mu}\phi\label{F}
\end{equation}
which has already been demonstrated in \cite{RPgal}. Thus we could relabel the derivative $\bar{D}_{\mu}\pi$, appearing in \ref{cov} as follows,
\begin{equation}
\bar{D}_{\mu}\pi=\partial_{\mu}(\phi+\pi)=\partial_{\mu}\pi{'}\label{d}
\end{equation}
This relabeling will show interesting implications in writing the Lagrangian of the Galileon field in curved backgrounds. To understand it better, we now introduce the inverse metric,
\begin{equation}
g^{\mu\nu}=\eta^{\alpha\beta}\Sigma_{\alpha}{}^{\mu}\Sigma_{\beta}{}^{\nu}\label{invmet}
\end{equation}
Using \ref{invmet} we can write the action for Galileon field given by \ref{la2} in curved background as,
\begin{equation}
S_{\rm g}=\int d^4x\sqrt{-g}\left[-\frac{1}{2}\left(g^{\rho\sigma}\partial_{\rho}\pi'\partial_{\sigma}\pi'
\right)+\alpha\Box\pi'\left(g^{\rho\sigma}\partial_{\rho}\pi'\partial
_{\sigma}\pi'\right)\right]\label{finalaction}
\end{equation}
This action shows minimal coupling of $\pi'$ to the curved background, under a particular St\"ueckelberg transformation $\pi\rightarrow\pi'=\pi+\phi$. At this stage it will be worthwhile to derive the transformation rules of $\pi$ and $\phi$ and hence that of $\pi'$. To start with it will be useful to consider the transformation properties of $\pi$ and $F_{\mu}$ \cite{RPgal},
\begin{align}
\delta \pi&=-\xi^\lambda\partial_\lambda \pi+ c + b_\mu x^\mu\notag\\\delta F_\mu & = -\xi^\lambda\partial_\lambda
F_\mu-\partial_\mu\xi^\lambda F_\lambda -\partial_\mu c - x^\nu\partial_\mu b_\nu\label{delF}
\end{align}
Using \ref{F} and the transformation property for $F_{\mu}$ as in \ref{delF} we get,
\begin{equation}
\delta \phi=-\xi^{\lambda}\partial_{\lambda}\phi-c - x^\nu b_\nu+\tilde{\phi}
\end{equation}
where we have used $b_{\mu}=\partial_{\mu}\tilde{\phi}$ as $b_{\mu}$ satisfies, $D_{\mu} b_{\nu}-D_{\nu} b_{\mu}=\partial_\mu b_\nu - \partial_\nu b_\mu=0$. The new field $\pi'$ and its derivative will thus transform as follows,
\begin{align}
\delta \pi'=&-\xi^{\lambda}\partial_{\lambda}\pi'+\tilde{\phi}\notag\\\delta (\partial_{\mu}\pi')=&-\partial_{\mu}(\xi^{\lambda}\partial_{\lambda}\pi')+\partial_{\mu}\tilde{\phi}\label{pi}
\end{align}
One can arrive at the flat spacetime limit of the above transformation rules in a trivial manner. Since in flat spacetime there is no reason for localization of the symmetry the additional gauge field $\phi$ must vanish and hence the field $\pi'$ should transform to $\pi$. This can also be obtained by treating the transformation parameters to be constant, such that a solution of $b_\mu = \partial_\mu\tilde{\phi}$ is $\tilde{\phi}= x_\nu b^\nu$. Thus in flat spacetime the field $\pi$ will transform as in \ref{delF}.

Hence in flat spacetime the theory takes the expected form with appropriate transformation properties for the Galileon field. To contrast this approach with those presented in earlier literatures we present the $\mathcal{L}_{3}$ term in the Lagrangian of \cite{Galilean_1,ref3,Galilean_2,Galilean_3,Galileon_2nd_order_eqm_1}, which reads
\begin{equation}
\bar{S}_{\rm g}=\int d^4x\sqrt{-g}~\Box\Pi\left(g^{\rho\sigma}\partial_{\rho}\Pi\partial
_{\sigma}\Pi\right)
\label{actioncomp}
\end{equation}
This brings out the crucial difference between our model and those presented in earlier literatures, since the field $\Pi$ is the full Galileon field unlike the field $\pi'$ appearing in \ref{finalaction} of our model. Further the transformation for $\Pi$ in \ref{actioncomp} becomes $\Pi\rightarrow \Pi+c$, which should again be contrasted with transformation property of $\pi'$ presented in \ref{pi} or for that matter with $\pi$ as in \ref{delF}.

One can immediately apply the above procedure to the other Galileon Lagrangians as well, namely $\mathcal{L}_{4}$ and $\mathcal{L}_{5}$. For example, if we consider the $\mathcal{L}_{4}$ term in the Galileon Lagrangian, then localization of Galileon symmetry would lead to the following expression for the corresponding action in curved spacetime,
\begin{align}
S_{\rm g}^{(4)}&=\int d^{4}x~\frac{1}{\Sigma}
\left(\eta ^{\gamma \delta}\bar{\mathcal{D}}_{\gamma}\pi \bar{\mathcal{D}}_{\delta}\pi \right) 
\left[\left(\eta ^{\alpha \beta}\mathcal{D}_{\alpha}\bar{\mathcal{D}}_{\beta}\pi \right)^{2}
-\eta ^{\alpha \gamma}\eta ^{\beta \delta}\mathcal{D}_{\alpha}\bar{\mathcal{D}}_{\beta}\pi \mathcal{D}_{\gamma}\bar{\mathcal{D}}_{\delta}\pi\right]
\nonumber
\\
&=\int d^{4}x~\sqrt{-g}\Big\{ \left(g^{\mu \nu}g^{\rho \lambda}\nabla _{\mu}\pi'\nabla _{\rho}\pi'\nabla _{\nu}\pi'\nabla _{\lambda}\pi'\right)R
-4\left(g^{\mu \nu}\nabla _{\mu}\pi' \nabla _{\nu}\pi' \right) 
\nonumber
\\
&~~~~~~~~~~~~~~~~~~~~~~~~~~~~~~~~~~~~~~~~~~~~~~~
\times \left[\left(\square \pi'\right)^{2}-\left(\nabla _{\mu}\nabla _{\nu}\pi'\nabla ^{\mu}\nabla ^{\nu}\pi' \right)\right]\Big\}
\end{align}
This again is structurally similar (though not identical) with the curved spacetime Galileon Lagrangian as derived in \cite{Galileon_2nd_order_eqm_1}. In consonance with the situation depicted earlier for the $\mathcal{L}_{3}$ term, for $\mathcal{L}_{4}$ as well the field appearing in the curved spacetime Galileon Lagrangian is $\pi'$ rather than the original Galileon field $\pi$. A similar result holds for $\mathcal{L}_{5}$ as well. Thus we have established that the Galileon field $\pi$ has full internal shift symmetry even in curved spacetime, while the dynamical part of it (namely, $\pi'$), appearing in the complete Galileon Lagrangian, will satisfy a reduced symmetry transformation. This is in sharp contrast with the earlier covariant Galileon models presented in the literature and hence provides a natural pathway to Galileon theories in curved spacetime respecting shift symmetry.
\section{Dynamics of the model}\label{Section_03}

One of the key motivations of the present study is to investigate the cosmological implications of the covariant Galileon, which is hitherto not discussed in detail in the existing literature. The action of the model includes the gravity sector, as well as the Galileon Lagrangian discussed in the previous section, which corresponds to
\begin{equation}
S = S_{\rm g} + S_{\rm EH} 
\end{equation}
Where $S_{\rm g}$ is the covariant Galileon action including $\mathcal{L}_{3}$ and minimally coupled with a background field and $S_{\rm EH}$ is the Einstein-Hilbert action for gravity, which is essentially the integral of Ricci scalar. For convenience the action for the Galileon field $S_{\rm g}$ has been reproduced here,
\begin{equation}
S_{\rm g}=\int d^4x\sqrt{-g}\left[-\frac{1}{2}\left(g^{\rho\sigma}\nabla_{\rho}\pi'\nabla_{\sigma}\pi'\right)+\alpha\Box \pi'\left(g^{\rho\sigma}\nabla_{\rho}\pi'\nabla_{\sigma}\pi'\right)\right]
\label{finalaction1}
\end{equation}
Here we have used the fact that the Galileon $\pi'$ transforms as a scalar field under general coordinate transformation. Given the action for the Galileon field, one can easily compute the energy-momentum tensor associated with it, defined  as,
\begin{equation}
T_{\mu\nu} = -\frac{2}{\sqrt{-g}}\frac{\delta S_{\rm g}}{\delta g^{\mu\nu}}\label{emt}
\end{equation}
Taking the variation of the matter action with respect to the metric $g^{\mu\nu}$, one ends up with
\begin{align}
\delta S_{\rm g}&=\int d^4x \sqrt{-g}\Big[\delta g^{\mu\nu} \Big\lbrace-\frac{1}{2}g_{\mu\nu}\left(-\frac{1}{2}\nabla\pi'.\nabla\pi'+\alpha\Box \pi'\nabla \pi'.\nabla \pi'\right)-\frac{1}{2}\nabla_{\mu}\pi'\nabla_{\nu}\pi'+\alpha\left(\nabla_{\mu}\nabla_{\nu}\pi'\right)\nabla \pi'.\nabla \pi' 
\nonumber
\\
& \qquad \qquad \qquad \quad \phantom{\frac{g}{2}} + \alpha\Box\pi'\left(\nabla_{\mu}\pi'\nabla_{\nu}\pi'\right)\Big\rbrace 
+\alpha g^{\mu\nu}\left(-\delta\Gamma^{\rho}
_{\mu\nu}\partial_{\rho}\pi'\right)\nabla \pi'.\nabla \pi' \Big]
\nonumber
\\
&=\int d^4x \sqrt{-g}~\delta g^{\mu\nu}\Big[\frac{1}{2}g_{\mu\nu}\Big\lbrace \frac{1}{2}\left(\nabla \pi'.\nabla \pi'\right)-\alpha \Box \pi'\left(\nabla \pi'.\nabla \pi'\right)\Big\rbrace +\alpha\left(\nabla_{\mu}\nabla_{\nu}\pi'\right)\nabla \pi'.\nabla \pi'
\nonumber
\\
&\qquad \qquad \qquad \qquad +(\alpha\Box\pi'-\frac{1}{2})\nabla_{\mu}\pi'\nabla_{\nu}\pi'+g_{\mu\nu}\frac{\alpha}{2}\nabla_{\rho}(\nabla \pi'.\nabla \pi' \nabla^{\rho}\pi')-\alpha\nabla_{\mu}(\nabla \pi'.\nabla \pi' \nabla_{\nu}\pi')\Big]\label{acvar}
\end{align}
where we have used the following result for variation of the metric compatible connection due to a variation of the metric,
\begin{equation}
\delta\Gamma^{\rho}_{\mu\nu}=\frac{1}{2}g^{\rho\sigma}\left(\nabla_{\nu}\delta g_{\mu\sigma}+\nabla_{\mu}\delta g_{\nu \sigma}-\nabla_{\sigma}\delta g_{\mu\nu}\right)\label{cv}
\end{equation}
Thus finally comparing \ref{acvar} with \ref{emt}, the energy momentum tensor for the Galileon field takes the following form,
\begin{align}
T_{\mu\nu}&=\alpha\Big[-g_{\mu\nu}\nabla_{\rho}(\nabla \pi'.\nabla \pi') \nabla^{\rho}\pi'+\nabla_{\mu}(\nabla \pi'.\nabla \pi') \nabla_{\nu}\pi'+\nabla_{\nu}(\nabla \pi'.\nabla \pi') \nabla_{\mu}\pi'-2\Box\pi'\nabla_{\mu}\pi'\nabla_{\nu}\pi'\Big]
\nonumber
\\
&+\nabla_{\mu}\pi'\nabla_{\nu}\pi'-\frac{1}{2}g_{\mu\nu}(\nabla\pi'.\nabla\pi')\label{stressfinal}
\end{align}
Note that the energy momentum tensor depends on at most second derivatives of the metric. Since the gravity action corresponds to the Einstein-Hilbert term alone, the Einstein's equations would become, $G_{\mu \nu}=8\pi G T_{\mu \nu}$, where $T_{\mu \nu}$ is the one given by \ref{stressfinal} and $G$ is the Newton's constant. 

If the above variation is due to a diffeomorphism, i.e., $\delta g^{\mu \nu}=\nabla ^{\mu}\xi ^{\nu}+\nabla ^{\nu}\xi ^{\mu}$, such that $\xi ^{\mu}$ vanishes at the boundary, one ends up with covariant conservation of $T^{\mu \nu}$, provided the Galileon field satisfies its own field equation \cite{Padmanabhan:2010zzb}. Thus for internal consistency of the theory the condition $\nabla _{\mu}T^{\mu}_{\nu}=0$ must lead to field equations for the Galileon field. Upon acting covariant derivative on the stress energy tensor given by \ref{stressfinal} one obtains
\begin{align}
\nabla^{\mu}T_{\mu\nu}&=\Box \pi'\nabla_{\nu}\pi'-\frac{1}{2}\nabla_{\nu}(\nabla\pi'.\nabla\pi')+\nabla_{\mu}\pi'\nabla^{\mu}\nabla_{\nu}\pi'
+\alpha\Big[\Box(\nabla\pi'.\nabla\pi')\nabla_{\nu}\pi'+\Box\pi'\nabla_{\nu}(\nabla\pi'.\nabla\pi')
\nonumber
\\
&-2(\Box\pi')^2\nabla_{\nu}\pi'-2\nabla_{\mu}\pi'\nabla^{\mu}\nabla
_{\nu}\pi'\Box\pi'\Big]
-\alpha\Big[2\nabla_{\mu}\pi'\nabla_{\nu}\pi'
\nabla^{\mu}(\Box\pi')+\nabla_{\nu}\nabla^{\rho}\pi'\nabla_{\rho}(\nabla\pi'.\nabla\pi')
\nonumber
\\
&-\nabla_{\mu}\pi'\nabla^{\mu}\nabla_{\nu}(\nabla\pi'.\nabla\pi')-\nabla_{\nu}\nabla^{\rho}\pi'\nabla_{\rho}(\nabla\pi'.\nabla\pi')\Big]
\nonumber
\\
&=\nabla_{\nu}\pi'\Big[\Box\pi'+\alpha\Box(\nabla\pi'.\nabla\pi')\Big]+\alpha\Big[\Box\pi'\nabla_{\nu}(\nabla\pi'.\nabla\pi')
\nonumber
\\
&-2\nabla_{\nu}\pi'(\Box\pi')^2-2\nabla_{\mu}\pi'\nabla^{\mu}\nabla_{\nu}
\pi'\Box\pi'-2\nabla_{\mu}\pi'\nabla_{\nu}\pi'\nabla^{\mu}(\Box\pi')\Big]
\end{align}
From the above expression it is not evident that the energy momentum tensor to be covariantly conserved. To see that the conservation actually holds, one must keep in mind that the $T_{\mu\nu}$ is conserved modulo the scalar field equation for $\pi'$ is satisfied. The field equation for $\pi'$ can be obtained from the variation of the action with respect to $\pi'$, leading to
\begin{align}
\delta S_{\rm g}&=\int d^4 x \sqrt{-g}\Big[-g^{\mu\nu}\nabla_{\mu}\pi'
\nabla_{\nu}\delta \pi'+2 \alpha\Box \pi'g^{\mu\nu}\nabla_{\mu}\pi'
\nabla_{\nu}\delta \pi'+\alpha\Big\lbrace (\nabla \pi'.\nabla \pi')\frac{1}{\sqrt{-g}}\partial_{\mu}(\sqrt{-g}g^{\mu\nu}\partial_{\nu}\delta \pi')\Big\rbrace\Big]
\nonumber
\\
&=(\textrm{total~div.})+\int d^4 x \sqrt{-g}\Big[\Box\pi'\delta\pi'-2\alpha\nabla^{\mu} (\Box \pi'\nabla_{\mu}\pi')\delta \pi'-\alpha\Big\lbrace \partial_{\mu}(\nabla \pi'.\nabla \pi')g^{\mu\nu}\partial_{\nu}\delta \pi')\Big\rbrace\Big]
\nonumber
\\
&=(\textrm{total~div.})+\int d^4 x \sqrt{-g}\Big[\Box\pi'-2\alpha\nabla^{\mu} (\Box \pi'\nabla_{\mu}\pi')+\alpha\frac{1}{\sqrt{-g}}\partial_{\nu}\Big\lbrace \sqrt{-g}g^{\mu\nu}\partial_{\mu}(\nabla \pi'.\nabla \pi')\Big\rbrace \Big]\delta \pi'
\end{align}
Thus the field equation for the Galileon field becomes,
\begin{equation}
\Box\pi'+\alpha \Big[\Box(\nabla \pi'.\nabla \pi')-2\nabla^{\mu}(\Box \pi'\nabla_{\mu}\pi')\Big]=0\label{eom}
\end{equation}
The above equations seems to violate the basic condition that field equation for Galileon field must have at most second derivatives. However this problem can be eliminated by introducing curvature tensor at the cost of commutator of covariant derivatives. In particular one can use the following result: $\nabla _{\alpha}\nabla _{\beta}\nabla _{\mu}\pi'-\nabla _{\mu}\nabla _{\alpha}\nabla _{\beta}\pi'=R_{\alpha \mu \beta \nu}\nabla ^{\nu}\pi'$ to finally obtain,
\begin{align}\label{Curv_Couple}
\Box \pi'+\alpha \Big[-2\left(\Box \pi'\right)^{2}+2R_{\mu \nu}\nabla ^{\mu}\pi'\nabla ^{\nu}\pi'+2\nabla _{\mu}\nabla _{\nu}\pi'\nabla ^{\mu}\nabla ^{\nu}\pi'\Big]=0
\end{align}
which has only second derivatives of the Galileon field and metric. Thus we have met our expectations, i.e., the field equation for the Galileon field is second order. Finally using the field equation in the original form as in \ref{eom} in the divergence of the energy momentum tensor we obtain,
\begin{align}
\nabla^{\mu}T_{\mu\nu}&=\alpha\Big[-2\nabla_{\nu}\pi'(\Box\pi')^2+2\nabla_{\nu}
\pi'\nabla^{\mu}(\nabla_{\mu}\pi'\Box\pi')+\Box\pi'\nabla_{\nu}(\nabla\pi'.\nabla\pi')
\nonumber
\\
&-2\nabla_{\mu}\pi'\nabla_{\nu}\pi'\nabla^{\mu}(\Box\pi')-2\nabla_{\mu}\pi'\nabla^{\mu}\nabla_{\nu}
\pi'\Box\pi'\Big]=0\label{cont}
\end{align}
Thus the stress energy tensor is indeed conserved as the Galileon field satisfies its field equation, proving internal consistency of our model. Hence we have the complete dynamics of the model at hand, the evolution of the Galileon field is being determined by its field equation, i.e., \ref{Curv_Couple}, which is second order but involves Ricci tensor as well. While the evolution of the metric is being determined by Einstein's equations (however also see \cite{Chakraborty:2014rga,Chakraborty:2015wma}) with the energy momentum tensor given in \ref{stressfinal}. In the next sections we will try to solve these system of equations analytically or numerically in the context of cosmological spacetime.
\section{Cosmological Application}\label{Section_04}

In this section we will study the cosmological implications of the covariant Galileon model introduced and elaborated above. We will adopt the metric describing a spacetime whose spatial section is maximally symmetric, known as the Friedmann-Robertson-Walker (FRW) spacetime because of the isotropy and homogeneity of the Universe. The metric compatible with the above symmetry requirements has the following structure,
\begin{equation}
ds^2=-dt^2+a(t)^2\left[\frac{dr^2}{1-\kappa r^2}+r^2(d\theta^2+\sin^2\theta d\phi^2)\right]\label{frwmetric}
\end{equation}  
where ($r, \theta, \phi$) are the comoving coordinates, $a(t)$ is known as scale factor and $\kappa$ is the constant representing the curvature of the three-space. Since from observations the universe at large scales appears spatially flat, we will only consider the case of zero curvature, i.e., $\kappa=0$. The Hubble parameter can be introduced as,
\begin{equation}
H(t)=\frac{\dot{a}(t)}{a(t)}\label{hubble}
\end{equation}
Given the above spacetime structure one can study the dynamics of the scale factor, which requires to solve the Einstein's equations. For this purpose one needs the energy momentum tensor for the Galileon field in a curved background, which we have evaluated in the previous section. To start with the computation of energy momentum tensor in FRW spacetime we need to find an expression for $\Box\pi'$ in terms of the Hubble parameter, keeping in mind that Galileon field is a function of cosmological time alone, $\pi'=\pi'(t)$, leading to,
\begin{align}
\Box\pi'=\frac{1}{a^3 r^2\sin\theta}\partial_{\mu}\left[a^3r^2\sin\theta ~g^{\mu\nu}\partial_{\nu}\pi'\right]=-\ddot{\pi}'-3H\dot{\pi}'\label{box}
\end{align}
Given this along with the result that $\nabla\pi'.\nabla\pi'=-\dot{\pi'}^2$ one can evaluate the stress energy tensor components using \ref{stressfinal} for the FRW spacetime resulting into,
\begin{align}
T^{\mu}_{\nu}&=\textrm{diag}\left[-\left(\frac{1}{2}\dot{\pi'}^2+6\alpha H\dot{\pi'}^3\right),\left(\frac{1}{2}\dot{\pi'}^2 -2\alpha\dot{\pi'}^2\ddot{\pi}'\right), \left(\frac{1}{2}\dot{\pi'}^2 -2\alpha\dot{\pi'}^2\ddot{\pi}'\right), \left(\frac{1}{2}\dot{\pi'}^2 -2\alpha\dot{\pi'}^2\ddot{\pi}'\right)\right]
\nonumber
\\
&\equiv(-\rho _{\rm g}, p_{\rm g}, p_{\rm g}, p_{\rm g})\label{stress}
\end{align}
Given the above diagonal structure of the energy momentum tensor one can immediately treat the Galileon field as a perfect fluid and can identify the density and pressure associated with it as,
\begin{align}
\rho_{\rm g}=\frac{1}{2}\dot{\pi'}^2+6\alpha H\dot{\pi'}^3;\qquad 
p_{g}=\frac{1}{2}\dot{\pi'}^2-2\alpha\dot{\pi'}^2\ddot{\pi}'\label{stress1}
\end{align}
Since one can represent the energy momentum tensor in the FRW backgrounds as a perfect fluid it immediately follows that the field equation for the Galileon field can be derived using the energy conservation equation $\dot{\rho}_{\rm g}+3H(\rho_{\rm g}+p_{\rm g})=0$, which from \ref{stress1} leads to,
\begin{align}
\dot{\pi}'(\ddot{\pi}'+3H\dot{\pi'})+6\alpha(\dot{H}\dot{\pi}'^3+2H\dot{\pi}'^2
\ddot{\pi}'+3H^2\dot{\pi}'^3)=0\label{xi}
\end{align}
This equation determines the time evolution of the Galileon field, but depends on the time evolution of the scale factor as well, through the Hubble parameter. Determining the time evolution of the scale factor $a(t)$ require the Einstein's field equations to be satisfied. Writing down Einstein's equations $G_{\mu\nu}=8\pi G T_{\mu\nu}$, yields the following Friedmann equation on spatially flat FRW backgrounds,
\begin{equation}
H^2=\frac{8\pi G}{3}\left(\frac{1}{2}\dot{\pi}'^2+6\alpha H\dot{\pi}'^3\right)\label{fried}
\end{equation}
The above equation being an algebraic one for the Hubble parameter, can be readily solved, leading to the following expression for $H(t)$,
\begin{align}
2H=6\alpha\beta\dot{\pi}'^3\pm \sqrt{36\alpha^2\beta^2\dot{\pi}'^6+2\beta\dot{\pi}'^2}\label{h}
\end{align}
where we have introduced $\beta=(8\pi G/3)$ for notational convenience. One can analyze this expression in two limiting situations, corresponding to $\alpha \rightarrow 0$ and $\alpha \rightarrow \infty$. In the $\alpha\rightarrow 0$ limit, the canonical kinetic term will dominate and as a consequences the Hubble parameter will have contribution from the kinetic term alone and has been well studied in the literature. We are interested in the opposite scenario with large value of $\alpha$. In large $\alpha$ limit the Hubble parameter reads, 
\begin{equation}
H=6\alpha\beta\dot{\pi}'^3=(16\pi G)\dot{\pi}'^3\label{hlarge}
\end{equation} 
In this limit the evolution equation for the Galileon field simplifies to,
\begin{align}
\dot{H}\dot{\pi'}^3+2H{\dot{\pi'}^2}\ddot{\pi'}+3H^2\dot{\pi'}^3 =0\label{eomstress}
\end{align}
Assuming $\dot{\pi}'\neq 0$ (i.e., $H\neq 0$) and using \ref{hlarge} and \ref{eomstress} we obtain the following differential equation satisfied by $\pi'$ alone,
\begin{equation}
\ddot{\pi'}+\frac{6}{5}(8\pi G)\dot{\pi'}^4=0\label{pid}
\end{equation}
This differential equation can be readily solved leading to the following time evolution of the Galileon field,
\begin{align}
\pi'(t)=B_2+\frac{1}{2\eta}[3\eta t-3B_1]^{2/3};~~\eta=\frac{6}{5}(8\pi G)\label{b}
\end{align}
where $B_{1}$ and $B_{2}$ corresponds to two arbitrary constants. Note that for large values of $t$ the Galileon field is real, which acted as a deciding criteria to pick out this particular solution. Given the solution for the Galileon field one can immediately obtain it's time derivative following \ref{b} as, $\dot{\pi'}=[3\eta t-3B_{1}]^{-1/3}$. Given the time derivative of the Galileon field we can immediately derive the Hubble parameter by using \ref{hlarge}, leading to,
\begin{align}
H &=\frac{5\eta}{3[3\eta t-3B_{1}]}
\end{align}
Before continuing further we will provide an explicit solution for the Galileon field. The constant $B_1$ appearing in the above expression can be determined using the fact that at present epoch $t=t_0$, $H=H_0$, resulting into,
\begin{equation}
B_1=\eta t_0-\frac{5\eta}{9H_0}\label{c1}
\end{equation}
Using this result for the arbitrary constant $B_{1}$ in the expression for the Galileon field $\pi'(t)$ we obtain,
\begin{equation}
\pi'(t)=B_2+\frac{1}{2\eta}\left[3\eta (t-t_0)+\frac{5\eta}{9H_0}\right]^{2/3}
\end{equation}
Finally, one can compute $\ddot{\pi}'$ having the expression: $\ddot{\pi'} =-\eta[3\eta t-3B_{1}]^{-4/3}$. Combining the expressions for the Hubble parameter along with $\dot{\pi}'(t)$ and $\ddot{\pi}'(t)$ in the large $\alpha$ limit the EoS parameter $\omega$ becomes,
\begin{equation}
\omega \equiv\frac{p}{\rho}=-\frac{\ddot{\pi}'}{3H\dot{\pi}'}=\frac{1}{5}
\end{equation}
Hence $\omega$ is positive for large $\alpha$. On the other hand without the interaction term (i.e., $\alpha=0$) the EoS parameter satisfies $\omega=1$ (see \ref{stress1}). We will now show that for $0\leq \alpha < \infty$, $\omega$ is always positive. If on the contrary $\omega$ becomes negative for some values of $\alpha=\alpha_{-}$, then there exists a point $\alpha=\alpha_{0}$ between $0$ and $\alpha_{-}$ where $\omega$ vanishes. But from (\ref{stress1}) and the definition of $\omega$, we can see that $p_g=0$ implies either $\dot{\pi}'=0$ or $\ddot{\pi}'=\frac{1}{4\alpha}$. $\dot{\pi}'$ cannot vanish everywhere. Also from \ref{pid} we find that if $\ddot{\pi}'=\frac{1}{4\alpha}$ then $(\dot{\pi}')^4$ becomes negative leading to another impossibility. So $\omega$ cannot vanish anywhere. It has to be always positive. Thus our model of covariant Galileon rules out late time cosmic acceleration, which demands $-1<\omega<-(1/3)$. 

Thus we have shown that within the premise of $\mathcal{L}_{3}$ alone it is not possible to get late time acceleration. A similar conclusion holds even when $\mathcal{L}_{4}$ and $\mathcal{L}_{5}$ terms are introduced in the covariant Galileon Lagrangian. To see the above assertion in some more detail it will be beneficial to first discuss how one may obtain late time acceleration in the context of Galileon theories. The standard paradigm is to break the Galileon symmetry in \emph{both} flat and curved spacetime by introducing arbitrary functions of $(\partial _{\mu}\pi'\partial ^{\mu}\pi')$. For example, the kinetic term is written as $G_{2}(X)$, while $\mathcal{L}_{3}$ becomes $G_{3}(X)\square \pi'$, where $X=-\partial _{\mu}\pi'\partial ^{\mu}\pi'$ with $G_{2}$ and $G_{3}$ being arbitrary functions of $X$. However these symmetry breaking terms have coupling coefficients which can be assumed to be small and hence the shift symmetry is weakly broken. The Lagrangian obtained with these 
modifications is referred to as weakly broken Galileon theories \cite{P}. The main difference between our approach with that of weakly broken Galileon theories is that ours preserve the full internal shift symmetry in curved spacetime, while the weakly broken Galileon theories break the shift symmetry not only in curved but also in flat spacetime. Breaking the shift symmetry ensures that one can have exact de Sitter solution, provided $\pi'(t)\sim t$ and 
\begin{align}
\partial _{X}G_{2}-3Z\partial _{X}G_{3}+6Z^{2}\left(\frac{1}{X}\partial _{X}G_{4}+2\partial _{X}^{2}G_{4}\right)
+Z^{3}\left(\frac{3}{X}\partial _{X}G_{5}+2\partial _{X}^{2}G_{5}\right)=0
\end{align}
where $G_{4}$ and $G_{5}$ are arbitrary functions associated with $\mathcal{L}_{4}$ and $\mathcal{L}_{5}$ respectively and $Z$ is the Hubble parameter. In the case of weakly broken Galileon theories one can always choose these arbitrary functions such that a de-Sitter solution does exist, which has implications in the inflationary paradigm or late time acceleration \cite{P}. However if we consider the scenario presented in this work, where the internal shift symmetry is maintained even in curved spacetime, it follows that the functions $G_{2}$ to $G_{5}$ are not arbitrary but have the following functional forms, $G_{2}=X/2$, $G_{3}=-X$, $G_{4}=X^{2}=G_{5}$. Thus one can substitute them in the above relation, leading to the following algebraic equation for Hubble parameter $(1/2)+3Z+36Z^{2}+10Z^{3}=0$, having a single real solution for $Z$, which is negative. This indicates a negative Hubble parameter, which is incompatible with the expansion of the universe. Thus unless the shift symmetry is weakly broken 
one can never have de-Sitter solution originating from curved spacetime Galileon Lagrangian. In particular, pertaining to the late time acceleration, it follows that the conditions like $\partial _{X}G_{4}+2X\partial _{X}^{2}G_{4}=0$ must be satisfied \cite{P}. It is clear that this is not possible with $G_{4}=X^{2}$. This conclusively shows that preserving the internal shift symmetry cannot lead to late time acceleration and is consistent with previous literatures.

We have already explained one particular root of obtaining late time acceleration, by introducing arbitrary functions of the canonical kinetic term and hence breaking the Galileon symmetry. There can be another possibility of arriving at late time acceleration by introducing an appropriate potential term. It is evident that the potential term is not invariant under the internal shift symmetry. Hence we have to break the shift symmetry while retaining the other property of the Galileon field, namely equation of motion should be second order, to get interesting cosmological consequences. However due to complicated nature of the field equations it is difficult to derive the potential from first principles, a tedious root would be to guess the form of the potential by inspection. Nevertheless, there exists an alternative and more reliable root to arrive at the potential, namely the reconstruction scheme \cite{Carloni:2010ph,Nojiri:2009xh,Nojiri:2006be,Nojiri:2009kx,Bhattacharya:2015wlz,Banerjee:2016hom}. In 
the following we will try to apply the reconstruction method, ultimately obtaining a potential leading to late time cosmic acceleration.
\section{Breaking the shift symmetry: Reconstruction method} 

We have seen in the previous section that if one respects shift symmetry of the Galileon field then the late time cosmic acceleration of our universe cannot be achieved. Hence in order to have accelerating solution of the universe at late times one has to break the Galileon shift symmetry. In principle one can break the symmetry in numerous possible ways by including symmetry breaking terms in the Lagrangian. However most of these terms will lead to higher order field equations for the Galileon field which will inevitably invite ghost modes to the theory. Thus the most economic choice would be to introduce a potential term for the Galileon field in the matter action. This will break the shift symmetry explicitly while keeping the field equations for the Galileon field second order and hence the shift symmetry is said to be \emph{weakly broken}. With this weakly broken Galileon symmetry the action of gravity plus matter field system takes the following form
\begin{align}
S=\int d^4 x\sqrt{-g}\left[\frac{R}{2\kappa^2}-\frac{1}{2}g^{\mu\nu}\partial_{\mu}\pi' \partial_{\nu}\pi'-V(\pi')+\alpha(g^{\mu\nu}\nabla_{\mu}\pi' \nabla_{\nu}\pi')\Box\pi'+L_{\rm matter}\right]\label{wbg}
\end{align}
where $\kappa^2=8\pi G$ is the four dimensional gravitational constant, $\alpha$ is a numerical coupling constant with dimension of $(\textrm{length})^{3}$ and $\pi'$ stands for the Galileon field. Note that we have included an additional piece $L_{\rm matter}$ containing all the additional matter fields present. As evident from the structure of the action given by \ref{wbg} addition of the potential term $V(\pi')$ breaks the internal Galileon symmetry. The variation of this action with respect to the metric can be carried out in a straightforward manner alike the situation in the previous section, ultimately yielding the following gravitational field equations,
\begin{align}
G_{\mu\nu}&=\kappa^2\Big[\nabla_{\mu}\pi'\nabla_{\nu}\pi'-g_{\mu\nu}\left\lbrace \frac{1}{2}g^{\alpha \beta}\nabla _{\alpha}\pi'\nabla_{\beta}\pi'+V(\pi')\right\rbrace 
+\alpha\Big\lbrace \nabla_{\nu}\pi'\nabla_{\mu}(g^{\alpha \beta}\nabla_{\alpha}\pi'\nabla_{\beta}\pi')
\nonumber
\\
&+\nabla_{\mu}\pi'
\nabla_{\nu}(g^{\alpha \beta}\nabla_{\alpha}\pi'\nabla_{\beta}\pi')
-2 \Box\pi'\nabla_{\mu}\pi'\nabla_{\nu}\pi'-{\frac{1}{2}}g_{\mu\nu}\nabla^{\rho}\pi'\nabla_{\rho}(g^{\alpha \beta}\nabla_{\alpha}\pi'\nabla_{\beta}\pi')\Big\rbrace+T_{\mu \nu}^{\rm matter}\Big]
\end{align}
where $T_{\mu \nu}^{\rm matter}$ stands for the energy momentum tensor associated with matter fields other than Galileon. On the other hand, the differential equation satisfied by the Galileon field $\pi'$ can be obtained by varying the action as in \ref{wbg} with respect to the Galileon field $\pi'$,
\begin{equation}
\Box\pi'+\alpha\Big\lbrace \Box(\nabla\pi'.\nabla\pi')-2\nabla^{\mu}(\Box\pi'\nabla_{\mu}\pi')\Big\rbrace-\frac{\partial V}{\partial\pi'}=0
\end{equation}
Note that the field equation for Galileon contains only second derivatives of the field and hence there exists no ghost modes in the theory. One can also easily prove that conservation of energy momentum tensor yields the same field equation for $\pi'$ even when the potential term is included, showing internal consistency of the method outlined above.

Having described the general features of this Galileon model with potential we will now turn our attention to the cosmological spacetime and shall depict that one can indeed arrive at a cosmological scenario with accelerated expansion of the universe. In this case as well the homogeneity and isotropy demands the Galileon field to be a function of time alone. This leads to the following expression for the energy momentum tensor associated with Galileon field,  
\begin{align}
T^{\mu~(g)}_{\nu}&=\textrm{diag}\left(-\rho_{\rm g},p_{\rm g},p_{\rm g},p_{\rm g}\right)
\\
\rho_{g}&=\frac{1}{2}\dot{\pi'}^2+6\alpha H\dot{\pi'}^3+V(\pi');\qquad 
p_{g}=\frac{1}{2}\dot{\pi'}^2-2\alpha\dot{\pi'}^2\ddot{\pi}'-V(\pi')\label{stressmod}
\end{align}
where an overdot denotes derivative with respect to cosmological time coordinate, $H=\dot{a}/a$ is the Hubble parameter and the subscript $g$ reminds that these contributions are coming from the Galileon field. Also note that unlike the previous scenario where pressure was always positive, here for small $\dot{\pi}'$ but large potential one can have negative pressure, an essential ingredient for the accelerating phase of the universe. Among the Friedmann equations the one expressing the Hubble parameter corresponds to 
\begin{equation}
H^2=\frac{\kappa^2}{3}\left(\rho_{m}+\rho_{g}\right)=\frac{\kappa^2}{3}\left[\rho _{m}+\frac{1}{2}\dot{\pi'}^2+6\alpha H\dot{\pi'}^3+V(\pi')\right]\label{f1}
\end{equation}
where $\rho_{m}$ denotes the energy density of the ordinary matter present in the spacetime, e.g., radiation, dark matter among others. Likewise the second independent  Einstein's equation is,
\begin{equation}
2(\dot{H}+H^2)+H^{2}=-\kappa^2 \left(p_{m}+p_{g}\right)=-\kappa^2\left[p_{m}+\frac{1}{2}\dot{\pi'}^2-2\alpha\dot{\pi'}^2\ddot{\pi}'-V(\pi')\right]\label{f2}
\end{equation}
where again $p_{m}$ denotes the pressure exerted by the matter field other than the Galileon. From \ref{f1} and \ref{f2} we get,
\begin{equation}
2\dot{H}=-\kappa^2\left[\left(\rho_{m}+p_{m}\right)+\dot{\pi'}^2+6\alpha H\dot{\pi'}^3+2\alpha\dot{\pi'}^2\ddot{\pi}'\right]\label{6}
\end{equation}
For consistency one can check that conservation of energy momentum tensor will lead to the continuity equation, which in this context, coincides exactly with the Einstein's equation presented above. This suggests that both the equations, namely \ref{f1} and \ref{6} are independent of each other. Another interesting point is that \ref{6} is independent of any potential term. 

One should now try to solve the above set of equations and obtain the Hubble parameter which in turn determines the scale factor of the universe. In principle this is a very difficult problem, since there may not exist any analytic solution for the above set of equations given an arbitrary potential. Similar situations have also been encountered in various other scenarios, in particular with higher curvature gravity. Since it is very difficult to solve those gravitational field equations one often invokes the reconstruction method, which we will briefly elaborate before borrowing this scheme for our case as well. 

In the reconstruction method one starts by assuming a particular choice of the scale factor, which explains the expansion history of the universe to a very good extent. Then using the scale factor in the gravitational field equations one can invert the procedure and obtain the corresponding form of the Lagrangian. In our case as well the form of the potential is unknown. Thus given a suitable scale factor describing evolution of our universe we can use the Friedmann equations to determine the structure of the potential. In a recent work using the reconstruction method the authors have tried to demonstrate the structure of the functions $G_{2}$, $G_{3}$ in weakly broken Galileon theories for either a bouncing or cyclic cosmology, here we apply it to the late time acceleration \cite{Chakraborty:2016ydo,Banerjee:2016hom}. 

To start with we consider the most simple scenario which corresponds to a constant Hubble parameter or equivalently an exponentially growing scale factor. In this case one can solve for the Galileon field analytically (neglecting the presence of additional matter fields) and the corresponding solution can be written as
\begin{align}
\pi'(t)&=-\frac{t}{3\alpha H} - \frac{c_{2}}{3H}e^{-3Ht}+c_1 \notag\\
&=-\frac{t}{3\alpha H}-\frac{c_2}{3H}\left(e^{-3Ht}-\frac{3H c_1}{c_2} \right)
\end{align}
where $c_{1}$ and $c_{2}$ are two arbitrary constants of integration which can be fixed by using appropriate boundary conditions. In particular one can possibly choose $\pi'(t=t_{1})=0=\dot{\pi}'(t=t_{1})$ for some time $t_{1}$. 
Let us now consider the limit of large $\alpha$, i.e., when the non-trivial Galileon term starts to dominate compared to the standard canonical kinetic term. In this limit is is clear that the term linear in time does not contribute and hence one obtains the Galileon field exclusively in terms of the exponential. Then using the other Friedmann equation one can obtain the potential for the Galileon field as a function of time. Finally eliminating the time in favour of the Galileon field one ultimately ends up with the potential as function of the Galileon field. In this particular situation one obtains, 
\begin{align}
\pi'(t)&=c_1 - \frac{c_2}{3H}e^{-3Ht}\notag\\
V(\pi')&=H^2 - 2 \alpha \kappa^2 H \left(c_2 e^{-3Ht}\right)^3= H^2 - 2 \alpha \kappa_n^2 H \left[3H\left(c_1 - \pi'\right)\right]^3 
\end{align}
This is the reconstruction potential in the large $\alpha$ limit. Note that the potential is cubic in the Galileon field. On the other hand for small $\alpha$, one would immediately obtain from the Friedmann equation with $\dot{H}=0$ that the Galileon field becomes time independent, i.e., $\dot{\pi}'=0$. This immediately results into, through the other Friedmann equation to $V(\pi')= \text{constant}$. This is the standard dark energy model with a scalar field having zero kinetic energy but constant potential energy, leading to equation of state parameter $\omega=-1$ and corresponds to the standard regime.

\begin{figure}
\begin{center}
\includegraphics[scale=0.6]{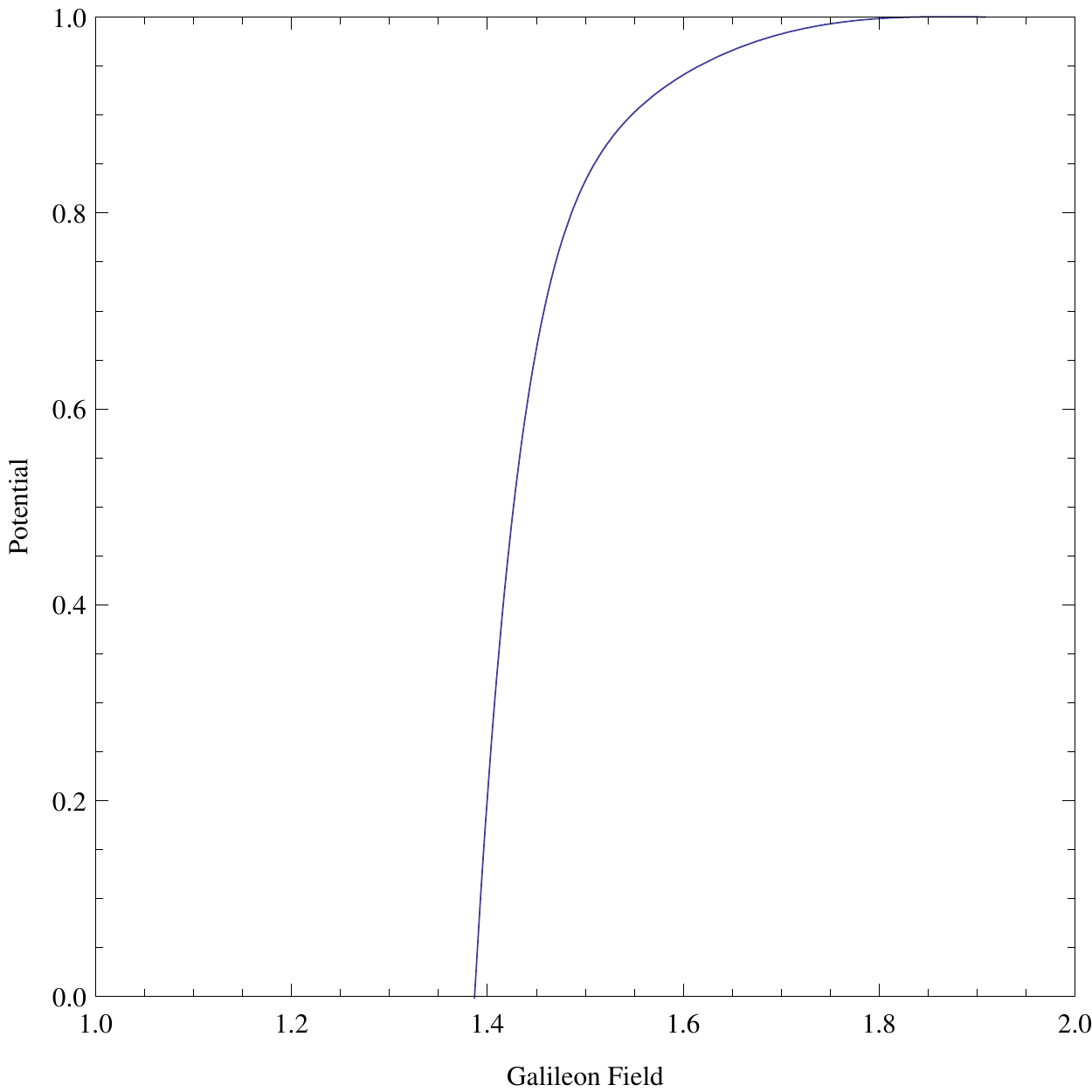}
\end{center}
\caption{The figure depicts the variation of the Galileon potential with the Galileon field which would result into late time cosmic acceleration. The potential has been constructed using the reconstruction method and solving the coupled differential equation for the scale factor and the Galileon field numerically.}
\label{fig_01}
\end{figure}

We will now consider a more general scenario by using an appropriate scale factor depicting the late time acceleration as well as matter dominating epoch in a single go. The corresponding scale factor depends on two arbitrary constants related to energy densities of the matter fields at the present epoch which reads, 
\begin{align}
\frac{a(t)}{a_0}=\left(\frac{c_{1}}{c_{2}}\right)^{\frac{1}{3}} \sinh^{\frac{2}{3}} \left[\frac{3}{2}\sqrt{c_{2}} H_0 t \right]\label{adot}
\end{align}
where $c_{1}$ and $c_{2}$ are two constants depending on the energy density of the matter fields at the present epoch and $H_{0}$ is the Hubble parameter at the present epoch. Note that for large values of the cosmological time parameter $t$ the scale factor do behave in an exponential manner. Given the scale factor one can compute its time derivative and hence the Hubble parameter and its derivative in a straightforward manner, leading to,
\begin{align}
H&=\frac{\dot{a}}{a}=\sqrt{c_{2}}H_0 \coth \left[\frac{3}{2}\sqrt{c_{2}} H_0 t \right]\label{hubble1}
\\
\dot{H}&=-\frac{3}{2}\sqrt{c_{2}} H_{0}^{2}~\textrm{cosech}^{2}\left(\frac{3}{2}\sqrt{c_{2}} H_0 t\right)  
\end{align}
Given the Hubble parameter and its derivative one can substitute them in the respective Friedmann equations. In particular one can use the equation for $\dot{H}$, which does not involve the potential and hence solve for the time dependence of the Galileon field. After obtaining $\pi'(t)$ one can substitute it back to the other Friedmann equation along with the expression for the Hubble parameter and hence obtain the corresponding solution for the potential as a function of time. Given both $\pi'(t)$ as well as $V(t)$ one can eliminate the cosmological time between these two and arrive at the corresponding potential for the Galileon field. In general the above procedure faces technical difficulties due to involved mathematical structure. However one can in principle apply standard numerical procedures to solve for the Galileon potential. Given the scale factor presented above one can use the numerical method and the corresponding potential is presented in \ref{fig_01}.  
It is clear that the potential saturates at larger values of the Galileon field and this is the kind of behaviour one expects from our earlier simple example as well. In that case also the potential was cubic and will show a very similar structure. Thus the numerical procedure for the more general situation leads to a potential of similar nature. 

Hence we have achieved our primary goal, i.e., to show that breaking of shift symmetry by addition of Galileon potential can indeed lead to late time accelerating solution for the universe in a cosmological scenario. While perusing this exercise we have also used the reconstruction scheme in the context of Galileon models in order to obtain the Galileon potential starting from the Friedmann equations and a viable cosmological scenario with late time acceleration.
\section{Conclusion}

In this paper we have successfully represented the covariant Galileon action derived in the metric formulation. By this process we have found a ghost-free Galileon model coupled to gravity in the curved background while respecting the internal shift symmetry. This is a new result based on an earlier work by two of us \cite{RPgal} where the flat space Galileon was lifted to the curved space by an appropriate localization of symmetries. This process of localization entailed the introduction of a new scalar $\phi$ (gauge field associated with shift symmetry). A St\"uckelberg type shift was achieved where the original flat space Galileon $\pi$ got replaced by $\pi'=\pi+\phi$. The transformation properties of the fields were also analyzed. A curved space generalization of the usual flat space shift symmetry was given. The important point of our construction was that the covariant Galileon was invariant under the curved space internal shift symmetry, apart from retaining its ghost free character. Interestingly, the original Galileon field $\pi$ enjoys the full internal shift symmetry. However the field appearing in the Galileon action in curved spacetime involves the field $\pi'$ having a reduced shift symmetry. We have also shown that our prescription leads to a covariant Galileon Lagrangian (see \ref{finalaction}) which is structurally similar (see \ref{actioncomp}) to that obtained in \cite{Galileon_2nd_order_eqm_1}. However the crucial difference with \cite{Galileon_2nd_order_eqm_1} is that our Lagrangian keeps the shift symmetry intact. This is a consequence of the transformation laws of basic fields in these two approaches.

It is well known in the literature that preserving the covariant description of Galileon retaining the internal shift symmetry is problematic. Indeed the covariant prescription was always done by breaking the shift symmetry but retaining the standard second order nature of the field equation. Generally, this is achieved by introducing arbitrary functions of the kinetic term in the Galileon Lagrangian and then choosing these arbitrary functions such that one can have late time acceleration. Taking a cue from the above, we have explicitly demonstrated that the curved spacetime Galileon Lagrangian satisfying shift symmetry, does \emph{not} lead to late time acceleration. This was shown by proving that the computed value of the EoS parameter $\omega$ lie within the range $\{(1/5),1\}$. The cosmological implications of the covariant Galileon theories with internal shift symmetry preserved were never explored earlier simply because such a model could not be constructed. We have successfully 
explored such a scenario.

Apart from introduction of arbitrary functions of the kinetic term, one can break the Galileon symmetry by introducing a potential for the Galileon field as well. Since in this context the potential is also a function of $\pi'$ alone, some portion of the original shift symmetry is still retained, e.g., $\pi\rightarrow \pi+c$ is still a symmetry of the potential term in curved spacetime. This is in sharp contrast with other approaches in this direction, where the introduction of potential term explicitly breaks the complete shift symmetry. Following this root, we have shown that introduction of a proper potential term, determined using reconstruction scheme, on FRW background does predict late time acceleration of universe. Hence we conclude, that only if one breaks the shift symmetry and introduces either an arbitrary function of the kinetic term or a potential term, the covariant Galileon model can predict late time acceleration. Our result is therefore compatible with earlier findings apart from providing 
new insights on the structure of the Galileon field in curved spacetime.
\section*{Acknowledgements}

Research of S.C. is supported by the SERB-NPDF grant (PDF/2016/001589) from SERB, Government of India. A part of this work was completed while S.C. was visiting Albert Einstein Institute in Golm, Germany and the author gratefully acknowledges the warm hospitality there.


\begin{thebibliography}{999}
 
\bibitem{obe19} S. Perlmutter {\it et al.} [Supernova Cosmology Project Collaboration], Astrophys. J. {\bf 517}, 565 (1999).

\bibitem{obe18} A. G. Riess et al. [Supernova Search Team Collaboration], Astron. J. 116, 1009 (1998).

\bibitem{obe3} K. N. Abazajian et al. [SDSS Collaboration], Astrophys. J. Suppl. 182, 543 (2009).

\bibitem{obe4} M. Kowalski et al. [Supernova Cosmology Project Collaboration], Astrophys. J. 686, 749 (2008)

\bibitem{obe1} E. Komatsu et al. [WMAP Collaboration], Astrophys. J. Suppl. 192, 18 (2011)

\bibitem{Dadhich:2010ca} N.~Dadhich, Pramana 77, 433 (2011) [arXiv:1006.1552 [gr-qc]].

\bibitem{Padmanabhan:2014nca} T.~Padmanabhan and H.~Padmanabhan, Int.\ J.\ Mod.\ Phys.\ D 23, no. 6, 1430011 (2014 [arXiv:1404.2284 [gr-qc]].

\bibitem{Padmanabhan:2017qvh} T.~Padmanabhan and H.~Padmanabhan, arXiv:1703.06144 [gr-qc].

\bibitem{DE} L. Amendola, S. Tsujikawa, {\it Dark Energy: Theory and Observations}, Cambridge University Press, 2010.

\bibitem{Paddy_1} T. Padmanabhan, Phys.Rept. 380, 235 (2003) [hep-th/0212290].

\bibitem{DE_Review_Sami} E. J. Copeland, M. Sami and S. Tsujikawa, Int. J. Mod.Phys. D 15, 1753 (2006) [hep-th/0603057].



\bibitem{obs_data_Phamtom_crossing} T. R. Choudhury and T. Padmanabhan, Astron. Astrophys. 429, 807 (2005).

\bibitem{Planck_1} P. A. R. Ade et al. [Planck Collaboration], arXiv:1303.5076 [astro-ph.CO].

\bibitem{Chakraborty:2014fia} S.~Chakraborty, S.~Pan and S.~Saha, Phys.\ Lett.\ B  738, 424 (2014) [arXiv:1411.0941 [gr-qc]].

\bibitem{N} A. Nicolis, R. Rattazzi and E. Trincherini, arXiv:0811.2197 [hep-th].

\bibitem{D} G. R. Dvali, G. Gabadadze and M. Porrati, Phys. Lett. B 485, 208 (2000).

\bibitem{LP} M. A. Luty, M. Porrati and R. Rattazzi, JHEP 0309, 029 (2003).
 
\bibitem{NR} A. Nicolis and R. Rattazzi, JHEP 0406, 059 (2004).

\bibitem{Galileon_2nd_order_eqm_1} C. Deffayet, G. Esposito-Farese, A. Vikman, Phys. Rev. D 79 (2009) 084003.

\bibitem{Galileon_2nd_order_eqm_2} C. Deffayet, S. Deser, G. Esposito-Farese, Phys. Rev. D 80 (2009) 064015.

\bibitem{Galileon_2nd_order_eqm_3} N. Chow, J. Khoury, Phys. Rev. D 80 (2009) 024037.

\bibitem{hamed} N. Arkani-Hamed, H. Georgi and M. D. Schwartz, Annals Phys. 305, 96 (2003).

\bibitem{Galilean_1} A. Nicolis, R. Rattazzi and E. Trincherini, Phys. Rev. D 79, 064036 (2009) [arXiv:0811.2197[hep-th]].

\bibitem{ref3} A. Ali, R. Gannouji and M. Sami, Phys. Rev. D 82, 103015 (2010).

\bibitem{Galilean_2} R. Gannouji, M. Sami, Phys.Rev.D 82, 024011 (2010) arXiv:1004.2808 [gr-qc].

\bibitem{Galilean_3} Tsutomu Kobayashi, Phys.Rev.D 81, 103533 (2010), arXiv:1003.3281 [astro-ph.CO].

\bibitem{P} D. Pirtskhalava, L. Santoni, E. Trincherini and F. Vernizzi, JCAP 1509 (2015) no.09, 007.
 
\bibitem{AASen11}  A. De. Felice and S. Tsujikawa, Phys. Rev. Lett. 105, 111301 (2010); S. A. Appleby and E. Linder,arXiv:1112.1981; E. Linder,arXiv:1201.5127.

\bibitem{AASen12} A. I. Vainshtein, Phys. Lett. B 39, 393 (1972).

\bibitem{Bhattacharya:2016naa} S.~Bhattacharya and S.~Chakraborty, Phys.\ Rev.\ D  95, 044037 (2017) [arXiv:1607.03693 [gr-qc]].

\bibitem{RPgal} R. Banerjee, P. Mukherjee, arXiv:1701.00971.

\bibitem{U}  R.~Utiyama, Phys.\ Rev.\   101, 1597 (1956).

\bibitem{K} T.~W.~B.~Kibble, J.\ Math.\ Phys.\   2, 212 (1961).

\bibitem{S} D. W. Sciama, in: Recent Developments in General Relativity, Festschrift for Infeld (Pergamon Press, Oxford) (1962) pp. 415.

\bibitem{W1} S. Weinberg, ``Gravitation and cosmology: principles and applications of the general theory of relativity,'' New York: Wiley (1972).

\bibitem{BH} M. Blagojevic and F.W. Hehl (eds.), ``Gauge Theories of Gravitation'', Imperial College Press, London (2013).

\bibitem{carroll} Sean Carroll, ``Space Time and Gravitation'', Pearson (2010).

\bibitem{MM1} R. Banerjee, A. Mitra, P. Mukherjee, Phys. Lett. B 737, 369-373 (2014).

\bibitem{MM2} R. Banerjee, A. Mitra, P. Mukherjee, Class.\ Quantum Grav. 32, 045010 (2015).

\bibitem{MM3} R. Banerjee, A. Mitra, P. Mukherjee, Phys. Rev. D 91, 084021 (2015).

\bibitem{BM4} R. Banerjee, P. Mukherjee, Phys. Rev. D 93, 085020 (2016).

\bibitem{MMM2} R. Banerjee, P. Mukherjee, Class. Quant. Grav. 22, 225013 (2016).

\bibitem{BGM} R.~Banerjee, S.~Gangopadhyay and P.~Mukherjee, arXiv:1604.08711 [hep-th].

\bibitem{Padmanabhan:2010zzb} T.~Padmanabhan, ``Gravitation: Foundations and frontiers,'' Cambridge, UK: Cambridge Univ. Press (2010).

\bibitem{Chakraborty:2014rga} S.~Chakraborty and T.~Padmanabhan, Phys.\ Rev.\ D  90, 124017 (2014) [arXiv:1408.4679 [gr-qc]].

\bibitem{Chakraborty:2015wma} S.~Chakraborty, JHEP 1508, 029 (2015) [arXiv:1505.07272 [gr-qc]].

\bibitem{Chakraborty:2016ydo} S.~Chakraborty and S.~SenGupta, Eur.\ Phys.\ J.\ C  76, 552 (2016) [arXiv:1604.05301 [gr-qc]].
 
\bibitem{Carloni:2010ph} S.~Carloni, R.~Goswami and P.~K.~S.~Dunsby, Class.\ Quant.\ Grav.\  {\bf 29}, 135012 (2012) [arXiv:1005.1840 [gr-qc]].

\bibitem{Nojiri:2009xh} S.~Nojiri, S.~D.~Odintsov, A.~Toporensky and P.~Tretyakov, Gen.\ Rel.\ Grav.\  {\bf 42}, 1997 (2010 [arXiv:0912.2488 [hep-th]].
  
\bibitem{Nojiri:2006be} S.~Nojiri and S.~D.~Odintsov, J.\ Phys.\ Conf.\ Ser.\  {\bf 66}, 012005 (2007) [hep-th/0611071].  

\bibitem{Nojiri:2009kx} S.~Nojiri, S.~D.~Odintsov and D.~Saez-Gomez, Phys.\ Lett.\ B {\bf 681}, 74 (2009) [arXiv:0908.1269 [hep-th]].

\bibitem{Bhattacharya:2015wlz} S.~Bhattacharya, P.~Mukherjee, A.~S.~Roy and A.~Saha, arXiv:1512.03902 [gr-qc].

\bibitem{Banerjee:2016hom} S.~Banerjee and E.~N.~Saridakis, Phys.\ Rev.\ D  95, 063523 (2017) [arXiv:1604.06932 [gr-qc]].

\end{thebibliography}
\end{document}